\documentclass[11pt]{revtex4-1}
\pdfoutput=1
\usepackage[latin9]{inputenc}
\usepackage{geometry}
\usepackage{prettyref}
\usepackage{textcomp}
\usepackage{amsmath}
\usepackage{amssymb}
\usepackage{stmaryrd}
\usepackage{graphicx}
\usepackage{esint}
\usepackage[unicode=true,pdfusetitle,
 bookmarks=true,bookmarksnumbered=false,bookmarksopen=false,
 breaklinks=false,pdfborder={0 0 1},backref=false,colorlinks=false]
 {hyperref}

\makeatletter
\usepackage{makeidx}\usepackage{amsfonts}

\makeatother

\begin{document}

\title{A Max-Sum algorithm for training discrete neural networks}

\author{Carlo Baldassi}

\affiliation{DISAT, Politecnico di Torino, Corso Duca Degli Abruzzi 24, 10129
Torino}

\affiliation{Human Genetics Foundation, Via Nizza 52, 10124 Torino}

\author{Alfredo Braunstein}

\affiliation{DISAT, Politecnico di Torino, Corso Duca Degli Abruzzi 24, 10129
Torino}

\affiliation{Human Genetics Foundation, Via Nizza 52, 10124 Torino}

\affiliation{Collegio Carlo Alberto, Via Real Collegio 1, Moncalieri}
\begin{abstract}
We present an efficient learning algorithm for the problem of training
neural networks with discrete synapses, a well-known hard (NP-complete)
discrete optimization problem. The algorithm is a variant of the so-called
Max-Sum (MS) algorithm. In particular, we show how, for bounded integer
weights with $q$ distinct states and independent concave \emph{a
priori} distribution (e.g. $l_{1}$ regularization), the algorithm's
time complexity can be made to scale as $O\left(N\log N\right)$ per
node update, thus putting it on par with alternative schemes, such
as Belief Propagation (BP), without resorting to approximations. Two
special cases are of particular interest: binary synapses $W\in\{-1,1\}$
and ternary synapses $W\in\{-1,0,1\}$ with $l_{0}$ regularization.
The algorithm we present performs as well as BP on binary perceptron
learning problems, and may be better suited to address the problem
on fully-connected two-layer networks, since inherent symmetries in
two layer networks are naturally broken using the MS approach.
\end{abstract}
\maketitle

\tableofcontents
\section{Introduction}

The problem of training an artificial, feed-forward neural network
in a supervised way is a well-known optimization problem, with many
applications in machine learning, inference etc. In general terms,
the problem consists in obtaining an assignment of ``synaptic weights''
(i.e.~the parameters of the model) such that the device realizes
a transfer function which achieves the smallest possible error rate
when tested on a given dataset of input-output examples. Time is usually
assumed to be discretized. In a single-layer network, the transfer
function is typically some non-linear function (e.g.~a sigmoid or
a step function) of the scalar product between a vector of inputs
and the vector of synaptic weights. In multi-layer networks, many
single-layer units operate in parallel on the same inputs, and their
outputs provide the input to other similar (with a varying degree
of similarity) units, until the last layer is reached.

The most popular and successful approaches to these kind of optimization
problems are typically variants of the gradient descent algorithm,
and in particular the back-propagation algorithm \cite{rumelhart1986learning}.
On single-layer networks with simple non-linearities in their output
functions these algorithms can even be shown to achieve optimal results
in linear time \cite{engel_statistical_2001}; on multi-layer networks
these algorithms suffer from the usual drawbacks of gradient descent
(mostly the presence of local minima, and slow convergence under some
circumstances).

On the other hand, gradient descent can only be applied to continuous
problems. If the synaptic weights are restricted to take only discrete
values, the abovementioned family of methods can not be applied; in
fact, it is known that even the simplest version of the problem (classification
using a single-layer network) becomes computationally hard (NP-complete)
in the worst-case scenario \cite{amaldi1994review,blum1992training}.
However, some theoretical properties of the networks, such as the
storage capacity (i.e.~the amount of information which can be effectively
stored in the device by setting the synaptic weights), are only slightly
worse in the case of discrete synapses, and other properties (e.g.~robustness
to noise and simplicity) would make them an attractive model for practical
applications. Indeed, some experimental results \cite{petersen1998all,o2005graded,bartol2015},
as well as arguments from theoretical studies and computer simulations
\cite{bhalla1999emergent,bialek2001stability,hayer2005molecular,miller2005stability},
suggest that long term information storage may be achieved by using
discrete --- rather than continuous --- synaptic states in biological
neural networks.

Therefore, the study of neural network models with discrete weights
is interesting both as a hard combinatorial optimization problem and
for its potential applications, in practical implementations as well
as for modeling biological networks. On the theoretical side, some
light has been shedded upon the origin of the computational hardness
in these kind of problems by the study of the space of the solutions
by means of methods derived from Statistical Physics approaches \cite{krauth_storage_1989,huang_origin_2014}:
in brief, most solutions are isolated, i.e.~far from each other,
and the energy landscape is riddled with local minima which tend to
trap purely local search methods, which thus show very poor performance.
On the application side, a family of heuristic algorithms, derived
from the cavity method, have been devised, which exhibit very good
performance on random instances, both in terms of solution time and
in terms of scaling with the size of the problem.

In particular, it was first shown in \cite{braunstein_learning_2006}
that a version of the Belief Propagation (BP) algorithm \cite{mezard_information_2009}
with the addition of a reinforcement term was able to efficiently
solve the problem of correctly classifying $\alpha N$ random input-output
associations using a single-layer network, or a tree-like two-layer
network, with $N$ synapses, up to a value of $\alpha$ close to the
theoretical upper bound. For the single-layer case, the theoretical
bound for binary synapses is $\alpha_{c}\simeq0.83$ \cite{krauth_storage_1989}, while
the algorithmic bound as estimated from extensive simulations up to
$N=10^{6}$ is $\alpha_{BP}\simeq0.74$. Two more algorithms, obtained
as crudely simplified versions of the reinforced BP, were later shown
\cite{baldassi_efficient_2007,baldassi_generalization_2009} to be
able to achieve very similar performances, despite being simpler and
working in an on-line fashion. The time complexity of all these algorithms
was measured to be of order $N\sqrt{\log N}$ per pattern; the BP
algorithm in particular achieves this performance thanks to a Gaussian
approximation which is valid at large $N$.

When considering multi-layer networks, the original BP approach of
\cite{braunstein_learning_2006} can only effectively deal with tree-like
network structures; fully-connected structures (such as those commonly
used in machine learning tasks) can not be addressed (at least not
straightforwardly) with this approach, due to strong correlations
arising from a permutation symmetry which emerges in the second layer.

In this paper, we present a new algorithm for addressing the problem
of supervised training of network with binary synapses. The algorithm
is a variant of the so-called Max-Sum algorithm (MS) \cite{mezard_information_2009}
with an additional reinforcement term (analogous to the reinforcement
term used in \cite{braunstein_learning_2006}). The MS algorithm is
a particular zero-temperature limit of the BP algorithm; but it should
be noted that this limit can be taken in different ways. In particular,
the BP approach in \cite{braunstein_learning_2006} was applied directly
at zero temperature as patterns had to be learned with no errors.
In the MS approach we present here, in addition to hard constraints
imposing that no errors are made on the training set, we add external
fields with a temperature that goes to zero in a second step. Random
small external fields also break the permutation symmetry for multi-layer
networks.

In the MS approach, the Gaussian approximation which is used in the
BP approach can not be used, and a full convolution needs to be computed
instead: this in principle would add a factor of $N^{2}$ to the time
complexity, but, as we shall show, the exploitation of the convexity properties of the problem
allows to simplify this computation, reducing the additional factor
to just $\log N$.

This reinforced MS algorithm has very similar performance to the reinforced
BP algorithm on single layer networks in terms of storage capacity
and of time required to perform each node update; however, the number
of updates required to reach a solution scales polynomially with $N$,
thus degrading the overall scaling. On fully-connected multi-layer
networks, the MS algorithm performs noticeably better than BP.

The rest of the paper is organized as follows: in \prettyref{sec:The-network-model}
we present the network model and the mathematical problem of learning.
In \prettyref{sec:The-Max-Sum-algorithm} we present the MS approach
for discrete weights. We show how the inherent equations can be solved
efficiently thanks to properties of the convolution of concave piecewise-linear
functions, and describe in complete detail the implementation for
binary weights. Finally, in \prettyref{sec:Numerical-results} we
show simulation results for the single and two-layer case.

\section{The network model\label{sec:The-network-model}}

We consider networks composed of one or more elementary ``building blocks''
(units), each one having a number of discrete weights and a binary transfer
function, which classify binary input vectors. Units can be arranged as a
composed function (in which the output from some units is considered as the
input of others) in various ways (also called architectures) that is able to
produce a classification output from each input vector.

We denote the input vectors as $\xi^{\mu}=\left\{ \xi_{i}^{\mu}\right\} _{i=1,\dots,N}\in\left\{ -1,+1\right\} ^{N}$
(where $\mu$ is a pattern index) and the weights as $W^{k}=\left\{ W_{i}^{k}\right\} _{i=1,\dots,N}$
(where $k$ is a unit index). In the following, the $W_{i}^{k}$ are
assumed to take $q$ evenly spaced values; we will then explicitly
consider the cases $q=2$ with $W_{i}^{k}\in\left\{ -1,1\right\} $
and $q=3$ with $W_{i}^{k}\in\left\{ -1,0,1\right\} $. The output
of the unit is given by:
\begin{equation}
\sigma^{k\mu}=\mathrm{sign}\left(\sum_{i=1}^{N}W_{i}^{k}\xi_{i}^{\mu}\right)
\end{equation}
with the convention that $\mathrm{sign}\left(0\right)=1$.

We will consider two cases: a single layer network and two-layer comitee machine. 
In single-layer networks, also called perceptrons \cite{rosenblatt1958perceptron},
there is a single unit, and therefore we will omit the index $k$.
Fully connected two-layer networks consist of $K$ units in the second
layer, each of which receives the same input vector $\xi^{\mu}$,
and the output function of the device is
\begin{equation}
\sigma^{\mu}=\mathrm{sign}\left(\sum_{k=1}^{K}\sigma^{k\mu}\right)\nonumber
\end{equation}

This kind of architecture is also called a committee or consensus
machine \cite{nilsson1965learning}. When $K=1$, this reduces to
the perceptron case. In a tree-like committee machine the input vectors
would not be shared among the units; rather, each unit would only
have access to a subset of the input vectors, without overlap between
the units. For a given $N$, the tree-like architectures are generally
less powerful (in terms of computational capabilities or storage capacity)
than the fully-connected ones, but are easier to train \cite{engel1992storage}.
Intermediate situations between these two extremes are also possible.
In fully-connected committee machines there is a permutation symmetry
in the indices $k$, since any two machines which only differ by a
permutation of the second layer's indices will produce the same output.

Throughout this paper we will consider supervised contexts, in which
each pattern $\mu$ has an associated desired output $\sigma_{D}^{\mu}$.

In \emph{classification} (or \emph{storage}) problems, $M=\alpha NK$
association pairs of input vectors $\xi^{\mu}$ and corresponding
desired outputs $\sigma_{D}^{\mu}$ are extracted from some probability
distribution, and the goal is to find a set of weights $\left\{ W^{k}\right\} _{k}$
such that $\forall\mu\in\left\{ 1,\dots,\alpha NK\right\} :\:\sigma^{\mu}=\sigma_{D}^{\mu}$.

In random \emph{generalization} problems, the input patterns $\xi^{\mu}$
are still extracted from some probability distribution, but the desired
outputs $\sigma_{D}^{\mu}$ are computed from some rule, usually from
a teacher device (\emph{teacher-student} problem). The goal then is
to learn the rule itself, i.e.~to achieve the lowest possible error
rate when presented with a pattern which was never seen during the
training phase. If the teacher's architecture is identical to that
of the student device, this can be achieved when the student's weights
match those of the teacher (up to a permutation of the units' indices
in the fully-connected case).

In the following, we will always address the problem of minimizing
the error function on the training patterns:
\begin{eqnarray}
E\left(\left\{ W^{k}\right\} _{k}\right) & = & \sum_{\mu=1}^{\alpha N}E_{\mu}\left(\left\{ W^{k}\right\} _{k}\right) - \sum_i \Gamma^k_i\left(W^k_i\right) \label{eq:Energy}\\
                                         & = & \sum_{\mu=1}^{\alpha N}\Theta\left(-\sigma_{D}^{\mu}\mathrm{sign}\left(\sum_{k=1}^{K}\mathrm{sign}\left(\sum_{i=1}^{N}W^{k}_i\xi_{i}^{\mu}\right)\right)\right) - \sum_i \Gamma^k_i\left(W^k_i\right)\nonumber 
\end{eqnarray}
where $\Theta\left(x\right)$ is the Heaviside step function $\Theta\left(x\right)=1$
if $x>0$ and $0$ otherwise. The term $\Gamma^k_i\left(W^k_{i}\right)$
has the role of an external field, and can be used e.g.~to implement a regularization
scheme; in the following, we will always assume it to be concave. For example we can implement $\l_1$ regularization by setting $\Gamma^k_i(W^k_i)=-\lambda |W^k_i|$ where $\lambda>0$ is a parameter.
 The first term of expression~\eqref{eq:Energy} therefore counts the number of misclassified patterns and the second one favours sparser solutions.

Throughout the paper, all random binary variables are assumed to be
extracted from an unbiased i.i.d.~distribution.

Under these conditions, it is known that in the limit of $N\gg1$
there are phase transitions at particular values of $\alpha$. For
single units (perceptrons) with binary $\pm1$ synapses, for the classification
problem, the minimum number of errors is typically $0$ up to $\alpha_{c}\simeq0.83$.
For the generalization problem, the number of devices which are compatible
with the training set is larger than $1$ up to $\alpha_{TS}\simeq1.245$,
after which the teacher perceptron becomes the only solution to the
problem.

\section{The Max-Sum algorithm\label{sec:The-Max-Sum-algorithm}}

Following \cite{braunstein_learning_2006}, we can represent the optimization
problem of finding the zeros of the first term of eq.~\eqref{eq:Energy} on a complete bipartite
factor graph. Starting from the single-layer case, the graph has $N$
vertices (variable nodes) representing the $W_{i}$ values and $\alpha N$
factor nodes representing the error terms $E_{\mu}\left(\mathbf{W}\right)$.

The standard MS equations for this graph involve two kind of messages
associated with each edge of the graph; we indicate with $\Phi_{\mu\to i}^{t}\left(W_{i}\right)$
the message directed from node $\mu$ to variable $i$ at time step
$t$, and with $\Psi_{i\to\mu}^{t}\left(W_{i}\right)$ the message
directed in the opposite direction.

These messages represent a certain zero-temperature limit of BP messages, but
have also a direct interpretation in terms of energy shifts of modified
systems. Disregarding an insubstantial additive constant, message $\Phi_{\mu\to
i}\left(W_{i}\right)$ represents the negated energy \eqref{eq:Energy}
restricted to solutions taking a specific value $W_i$ for variable $i$, on a
modified system in which the energy depends on $W_i$ only through the factor
node $\mu$, i.e.~in which all terms $W_i\xi_i^\nu$ for all $\nu\ne\mu$ are
removed from the energy expression \eqref{eq:Energy}.  Similarly, message
$\Psi_{i\to\mu}^{t}\left(W_{i}\right)$ represents an analogous negated energy
on a modified system in which the term $E_\mu$ is removed.  For factor graphs
that are acyclic, the MS equations can be thought of as the dynamic programming
exact solution. In our case, the factor graph, being complete bipartite, is far
from being acyclic and the equations are only approximate. For BP, the
approximation is equivalent to the one of the Thouless-Anderson-Palmers
equations \cite{kabashima_cdma_2003} and is expected to be exact in the
single-layer case below the critical capacity \cite{krauth_storage_1989}. For a
complete description of the MS equations and algorithm, see
\cite{mezard_information_2009}.

The MS equations \cite{mezard_information_2009} for energy \eqref{eq:Energy} are:
\begin{eqnarray}
\Phi_{\mu\to i}^{t+1}\left(W_{i}\right) & = & \max_{\left\{ W_{j}\right\} _{j\ne i}:\,E_{\mu}\left(\mathbf{W}\right)=0}\left(\sum_{j\ne i}\Psi_{j\to\mu}^{t}\left(W_{j}\right)\right)-Z_{\mu\to i}^{t+1}\label{eq:MS_CavPhi}\\
\Psi_{i\to\mu}^{t}\left(W_{i}\right) & = & \Gamma_{i}\left(W_{i}\right)+\sum_{\mu^{\prime}\ne\mu}\Phi_{\mu^{\prime}\to i}^{t}\left(W_{i}\right)-Z_{i\to\mu}^{t}\label{eq:MS_CavPsi}
\end{eqnarray}
where $Z_{\mu\to i}^{t}$ and $Z_{i\to\mu}^{t}$ are normalization
scalars that ensure $\sum_{W_{i}}\Psi_{i\to\mu}^{t}\left(W_{i}\right)=\sum_{W_{i}}\Psi_{\mu\to i}^{t}\left(W_{i}\right)=0$
and can be computed after the rest of the RHS.
At any given time $t$, we can compute the single-site quantities
\begin{equation}
\Psi_{i}^{t}\left(W_{i}\right)=\Gamma_{i}\left(W_{i}\right)+\sum_{\mu}\Phi_{\mu\to i}^{t}\left(W_{i}\right)-Z_{i}^{t}\label{eq:MS_Psi}
\end{equation}
and use them to produce an assignment of the $\mathbf{W}$'s:
\begin{equation}
W_{i}^{t}=\mathrm{argmax}_{W_{i}}\Psi_{i}^{t}\left(W_{i}\right)\label{eq:MS_Wt}
\end{equation}

The standard MS procedure thus consists in initializing the messages
$\Psi_{i\to\mu}^{0}\left(W_{i}\right)$, iterating eqs.~\eqref{eq:MS_CavPhi}
and \eqref{eq:MS_CavPsi} and, at each time step $t$, computing a
vector $W^{t}$ according to eqs.~\eqref{eq:MS_Psi} and \eqref{eq:MS_Wt}
until either $E\left(W^{t}\right)=0$ (in the absence of prior terms, i.e. when $\lambda=0$), or the messages converge to
a fixed point, or some maximum iteration limit is reached.

Strictly speaking, standard MS is only guaranteed to reach a fixed
point if the factor graph is acyclic, which is clearly not the case
here. Furthermore, if the problem has more than one solution (ground
state), the assignment in eq.~\eqref{eq:MS_Wt} would not yield a solution
even in the acyclic case. In order to (heuristically) overcome these
problems, we add a time-dependent reinforcement term to eqs.~\eqref{eq:MS_CavPsi}
and \eqref{eq:MS_Psi}, analogously to what is done for BP \cite{braunstein_learning_2006}:
\begin{eqnarray}
\Psi_{i\to\mu}^{t}\left(W_{i}\right) & = & r\,t\,\Psi_{i}^{t-1}\left(W_{i}\right)+\Gamma_{i}\left(W_{i}\right)+\sum_{\mu^{\prime}\ne\mu}\Phi_{\mu^{\prime}\to i}^{t}\left(W_{i}\right)-Z_{i\to\mu}^{t}\label{eq:RMS_CavPsi}\\
\Psi_{i}^{t}\left(W_{i}\right) & = & r\,t\,\Psi_{i}^{t-1}\left(W_{i}\right)+\Gamma_{i}\left(W_{i}\right)+\sum_{\mu^{\prime}}\Phi_{\mu^{\prime}\to i}^{t}\left(W_{i}\right)-Z_{i}^{t}\label{eq:RMS_Psi}
\end{eqnarray}
where $r>0$ controls the reinforcement speed. This reinforcement term in the case of standard BP implements a sort of ``soft decimation'' process, in which single variable marginals are iteratively pushed to concentrate on a single value. For the case of MS, this process is useful to aid convergence: on a system in which the MS equations do not converge, the computed MS local fields still give some information about the ground states and can be used to 
iteratively ``bootstrap'' the system into one with very large external fields, i.e. fully polarized on a single configuration \cite{bailly-bechet_finding_2011}. The addition of this term introduces a dependence on the initial condition.
Experimentally, by reducing $r$ this dependence can be made arbitrarily small, leading to more accurate results (see Sec.~\ref{sec:Numerical-results}), at the cost of increasing
the number of steps required for convergence; our tests show that the
convergence time scales as $r^{-1}$.

Furthermore, in order to break symmetries between competing configurations,
we add a small symmetry-breaking concave noise $\Gamma'_i\left(W_i\right)\ll 1$ to the external fields 
$\Gamma_{i}\left(W_{i}\right)$;
this, together with the addition of the reinforcement term, is sufficient
to ensure --- for all practical purposes --- that the $\mathrm{argmax}$
in \eqref{eq:MS_Wt} is unique at every step of the iteration.

\subsection{Max Convolution}

While Eq. \eqref{eq:RMS_CavPsi} can be efficiently computed in a
straightforward way, the first term of Eq. \eqref{eq:MS_CavPhi} involves
a maximum over an exponentially large set. The computation of Eq. \eqref{eq:MS_CavPhi} can be rendered tractable by adding $0=\max_{\Delta}L\left(\Delta,\sum_{j\neq i}\xi_{j}^{\mu}W_{j}\right)$ where $L\left(x,y\right)$ is equal to 0 if $x=y$ and $-\infty$ otherwise, which leads to the following transformations:

\begin{eqnarray}
\Phi_{\mu\to i}^{t+1}\left(W_{i}\right)+Z_{\mu\to i}^{t+1} & = & \max_{\left\{ W_{j}\right\} _{j\ne i}:\,E_{\mu}\left(\mathbf{W}\right)=0}\left(\sum_{j\ne i}\Psi_{j\to\mu}^{t}\left(W_{j}\right)\right)\nonumber\\
 & = & \max_{\left\{ W_{j}\right\} _{j\ne i}:\,E_{\mu}\left(\mathbf{W}\right)=0}\left\{ \max_{\Delta}L\left(\Delta,\sum_{j\neq i}\xi_{j}^{\mu}W_{j}\right)+\sum_{j\ne i}\Psi_{j\to\mu}^{t}\left(W_{j}\right)\right\}\nonumber \\
 & = & \max_{\Delta:\sigma^{\mu}_{D}\left(\Delta+\xi_{i}^{\mu}W_{i}\right)\geq0}\left\{ \max_{\left\{ W_{j}\right\} _{j\ne i}:\sum_{j\neq i}\xi_{j}^{\mu}W_{j}=\Delta}\sum_{j\ne i}\Psi_{j\to\mu}^{t}\left(W_{j}\right)\right\}\nonumber \\
 & = & \max_{\Delta:\sigma^{\mu}_{D}\left(\Delta+\xi_{i}^{\mu}W_{i}\right)\geq0}\mathcal{F}_{\mu\to i}^{t}\left(\Delta\right)\label{eq:message},
\end{eqnarray}
where in the last step above $\mathcal{F}_{\mu\to i}^{t}\left(\Delta\right)$
is defined as:
\begin{eqnarray}
\mathcal{F}_{\mu\to i}^{t}\left(\Delta\right) & = & \max_{\left\{ S_{j}\right\} _{j\ne i}:\,\sum_{j\neq i}S_{j}=\Delta}\left(\sum_{j\ne i}\Psi_{j\to\mu}^{t}\left(S_{j}\xi_{j}^{\mu}\right)\right).\label{eq:convolution}
\end{eqnarray}

The right-hand side of \eqref{eq:convolution} is usually called
a ``Max-Convolution'' of the functions $f_{j}\left(S_{j}\right)=\Psi_{j\to\mu}\left(S_{j}\xi_{j}^{\mu}\right)$
for $j\neq i$, and is analogous to the standard convolution but with
operations $\left(\max,+\right)$ substituting the usual $\left(+,\times\right)$.
As standard convolution, the operation is associative, which allows
to compute the convolution of the $N-1$ functions in a recursive
way. As the convolution of two functions with discrete domains $\left\{ 0,\dots,q_{1}\right\} $
and $\left\{ 0,\dots,q_{2}\right\} $ respectively can be computed
in $q_{1}q_{2}$ operations and has domain in $\left\{ 0,\dots,q_{1}+q_{2}\right\} $,
it follows that \eqref{eq:convolution} can be computed in $O\left(N^{2}\right)$
operations. In principle, this computation must be performed $N$
times for each pattern $\mu$ to compute all $\Phi_{\mu\to i}$ messages,
in a total of time $O\left(N^{3}\right)$. 

A technique like the one described in
\cite{braunstein_estimating_2008,braunstein_efficient_2011,braunstein_efficient_2009}
can be employed to reduce this by a factor $N$, coming back again to
$O\left(N^{2}\right)$ operations per pattern update, as follows. Precomputing
the partial convolutions $L_n$ and $R_n$ of $f_1,\dots,f_n$ and $f_n,\dots,f_N$
(respectively) for every $n$ in $1,\dots,N$ can be done using $N^2$ operations
in total; then the convolution of $\left\{f_j\right\}_{j\neq i}$ can be
computed as the convolution of $L_{i-1}$ and $R_{i+1}$. Computing this
convolution would require $\mathcal{O}(N^2)$ operations but fortunately this
will not be needed; the computation of \eqref{eq:message} can proceed as:
\begin{eqnarray}
  \Phi_{\mu\to i}^{t+1}\left(W_{i}\right)+Z_{\mu\to i}^{t+1} & = & \max_{\Delta:\sigma_{D}^{\mu}\left(\Delta+\xi_{i}^{\mu}W_{i}\right)\geq0}\left\{\max_{z}L_{i-1}\left(z\right)+R_{i+1}\left(\Delta-z\right)\right\}\nonumber\\
& = & \max_{z}\left\{L_{i-1}\left(z\right)+\max_{\Delta:\sigma_{D}^{\mu}\left(\Delta+\xi_{i}^{\mu}W_{i}\right)\geq0}R_{i+1}\left(\Delta-z\right)\right\}\nonumber\\
& = & \max_{z}\left\{L_{i-1}\left(z\right)+R_{i+1}^{\sigma_{D}^{\mu}}\left(z+\xi_{i}^{\mu}W_{i}\right)\right\}\label{eq:RR}
\end{eqnarray}
where we defined
$R_{i}^{\sigma}\left(x\right)=\max_{\Delta:\sigma\left(\Delta+x\right)\ge 0}R_{i}\left(\Delta\right)$.
As the vectors $R_{i}^{\sigma}$ can be
pre-computed recursively in a total time of $\mathcal{O}\left(N^{2}\right)$
and \eqref{eq:RR} requires time $\mathcal{O}\left(N\right)$, we obtain a grand
total $\mathcal{O}\left(N^{2}\right)$ operations per pattern update, or
$\mathcal{O}(M N^2)$ per iteration. Unfortunately, this scaling is normally
still too slow to be of practical use for moderately large values of $N$ and
we will thus proceed differently by exploiting convexity properties. However, note
that the above scaling is still the best we can achieve for the general case
in which regularization terms are not concave.

At variance with standard discrete convolution,
in general Max-Convolution does not have an analogous to the Fast Fourier Transform, that would allow a reduction of the computation
time of a convolution of functions with $\mathcal{O}(N)$ values
from $N^{2}$ to $N\log N$. Nevertheless, for
concave functions the convolution can be computed efficiently, as
we will show below. Note that for this class of functions, an operation
that is analogous to the Fast Fourier Transform is the Legendre-Fenchel
transform \cite{boudec_network_2001}, though it will be simpler to
work with the convolution directly in the original function space.

\begin{figure}

\begin{centering}
\includegraphics[width=0.4\columnwidth]{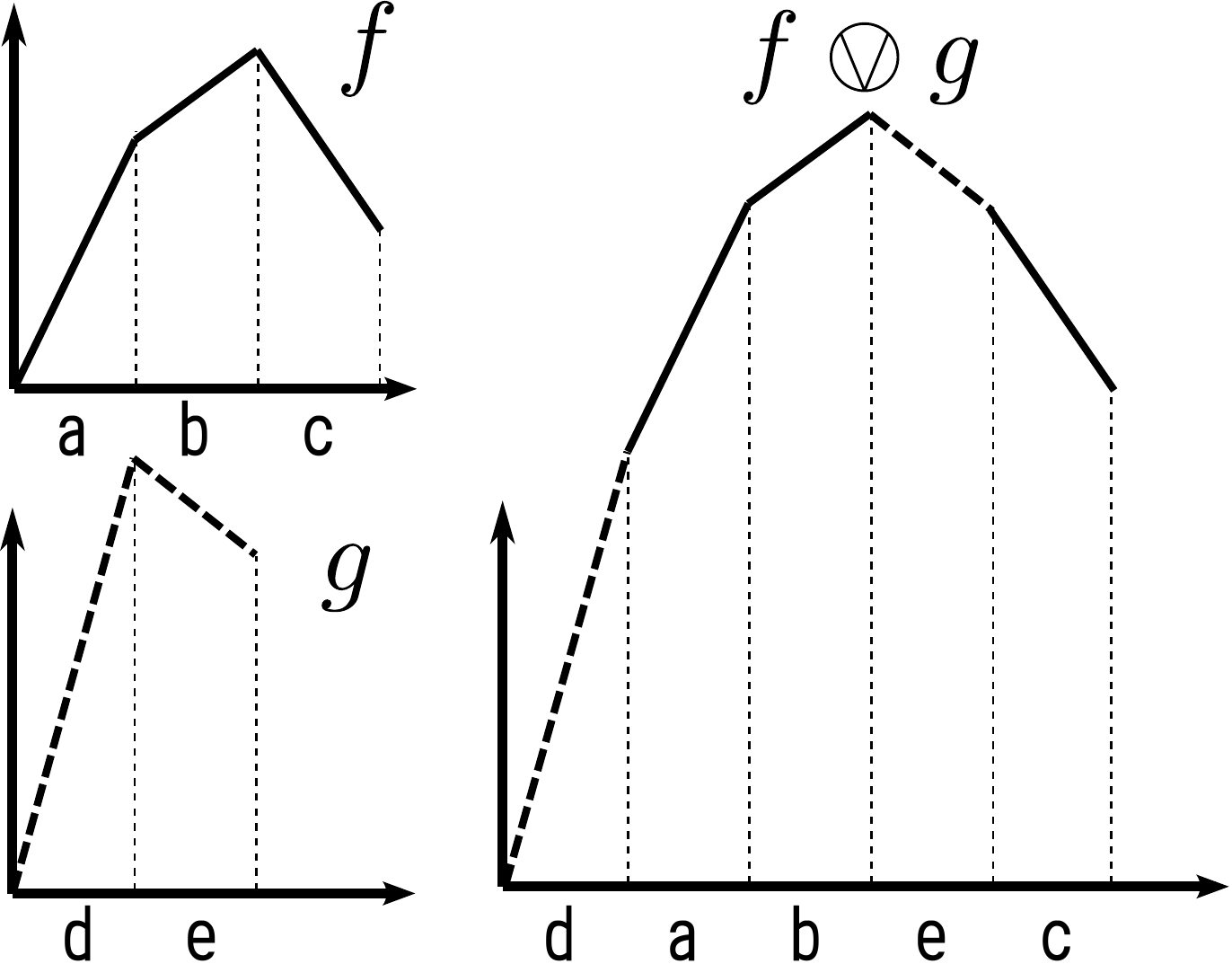}
\par\end{centering}

\protect\caption{\label{fig:convo}Sketch of the discrete max-convolution $f\ovee g$
of piecewise-linear concave functions $f,g$. The result is simply
obtained by sorting the pieces of the two functions in descending
order of slope.}

\end{figure}

First, let us recall well-known results about max-convolution of concave
piecewise-linear functions in the family $\mathcal{C}=\left\{ f:\mathbb{R}_{\geq0}\to\mathbb{R}\cup\left\{ -\infty\right\} \right\} $
\cite{boudec_network_2001}. First, the max-convolution $f_{12}\stackrel{def}{=}f_{1}\ovee f_{2}$
of $f_{1},f_{2}\in\mathcal{C}$ belongs to $\mathcal{C}$. Moreover,
$f_{1}\ovee f_{2}$ can be built in an efficient way from $f_{1}$
and $f_{2}$. Start with $x_{1}^{12}\stackrel{def}{=}\inf\left\{ x:\left(f_{1}\ovee f_{2}\right)\left(x\right)>-\infty\right\} $,
which is easily computed as $x_{1}^{12}=x_{1}^{1}+x_{1}^{2}$ with
$x_{1}^{i}=\inf\left\{ x:f_{i}\left(x\right)>-\infty\right\} $. Moreover,
$\left(f_{1}\ovee f_{2}\right)\left(x_{1}^{12}\right)=f_{1}\left(x_{1}^{1}\right)+f_{2}\left(x_{1}^{2}\right)$.
Then, order the set of linear pieces from $f_{1}$ and $f_{2}$ in
decreasing order of slope and place them in order, starting from $\left(x_{1}^{12},f\left(x_{1}^{12}\right)\right)$
to form a piecewise-linear continuous function. The method is sketched
in Fig.~\ref{fig:convo}. In symbols, let us write each concave piecewise-linear
function $f_{i}\left(x\right)$ when $x\ge x_{1}^{i}$ as: 
\[
f_{i}\left(x\right)=\sum_{j=1}^{k_{i}}\left(x-x_{j}^{i}\right)_{+}a_{j}^{i}+f_{i}\left(x_{1}^{i}\right)\quad i=1,2
\]
with $a_{j}^{i}\in[0,\infty]$ for $j=2,\dots,k_{i}$ and $x_{1}^{i}<x_{2}^{i}<\cdots<x_{k_{i}}^{i}$.
Here we used the notation\footnote{We allow $a_{k_{i}}^{i}=-\infty$ , in which case we conventionally
define $a_{k_{i}}^{i}y_{+}=0$ if $y\le0$.} $y_{+}=\frac{1}{2}\left(\left|y\right|+y\right)$. This function is concave, as for $x\in\left[x_{j}^{i},x_{j+1}^{i}\right]$
the slope is $b_{j}^{i}=\sum_{k=1}^{j}a_{k}^{i}$ that is clearly
decreasing with $j$. To compute the convolution of $f_{1}$ and $f_{2}$,
just order the slopes $b_{j}^{i};$ i.e.~compute a one to one map
$\pi:\left(i,j\right)\mapsto c$ from couples $i\in\left\{ 1,2\right\} $,$1\leq j\leq k_{i}$
to integers $1\leq c\leq k_{1}+k_{2}$ such that $\pi\left(i,j\right)<\pi\left(i^{\prime},j^{\prime}\right)$
implies $b_{j}^{i}>b_{j^{\prime}}^{i^{\prime}}$. The max convolution
for $x\ge x_{1}^{12}$ is still concave and piecewise-linear, and
thus it can be written as:
\[
\left(f_{1}\ovee f_{2}\right)\left(x\right)=\sum_{c=1}^{k_{12}}\left(x-x_{c}^{12}\right)_{+}a_{c}^{12}+\left(f_{1}\ovee f_{2}\right)\left(x_{1}^{12}\right)
\]
where $k_{12}=k_{1}+k_{2}-1$. For each $c$ we can retrieve $\left(i\left(c\right),j\left(c\right)\right)=\pi^{-1}\left(c\right)$;
with this, the parameters of the convolution are $a_{1}^{12}=b_{j\left(1\right)}^{i\left(1\right)}$,
$a_{c+1}^{12}=b_{j\left(c+1\right)}^{i\left(c+1\right)}-b_{j\left(c\right)}^{i\left(c\right)}$
and $x_{c+1}^{12}=x_{1}^{12}+x_{j\left(c\right)+1}^{i\left(c\right)}-x_{j\left(c\right)}^{i\left(c\right)}$.

For more details about the max-convolution of piecewise-linear concave
functions, see e.g. \cite[Part II]{boudec_network_2001}.

We now consider the case of functions defined on a discrete domain.
Let $f,g$ be concave discrete functions in 
\[
\mathcal{D=}\left\{ f:\left\{ 0,\dots,q-1\right\} \to\mathbb{R}\cup\left\{ -\infty\right\} \right\} 
\]
We will define the continuous extension $\hat{f}\in\mathcal{C}$ as
the piecewise-linear interpolation of $f$, with value $-\infty$
for arguments in $\left(-\infty,0\right)\cup\left(q-1,\infty\right)$.
This can be written as:
\[
\hat{f}\left(x\right)=\sum_{j=1}^{q}\left(x-j+1\right)_{+}a_{j}+f\left(0\right)
\]
with $a_{1}=f\left(1\right)-f\left(0\right)$, $a_{j+1}=f\left(j+1\right)-2f\left(j\right)+f(j-1)$
(implying $a_{q}=-\infty$). It is easy to see that $h=\hat{f}\ovee\hat{g}$
coincides with the discrete convolution of $f$ and $g$ in its (discrete)
domain; the reason is simply that $h$ is also piecewise-linear, with
kinks only in discrete values ,$\left\{ 0,\dots2\left(q-1\right)\right\} $.

When computing the convolution of $N$ functions $f_{1},\dots,f_{N}$
with domain in $\left\{ 0,\dots,q-1\right\} $, one can clearly order
all slopes together in one initial pass, using $Nq\log\left(Nq\right)$
comparisons. It is also easy to see that, if one has the full convolution,
it is simple to compute a ``cavity'' convolution in which one function
$f_{i}$ is omitted, in $O\left(q\right)$ time: this is achieved
by simply removing the $q$ slopes of $f_{i}$ from the full convolution.

In order to apply this to eq.~\eqref{eq:convolution} the only remaining
step is a trivial affine mapping of the functions arguments $W_{j}\xi_{j}^{\mu}$
on a domain $\left\{ A+tB:t\in\left\{ 0,\dots,q-1\right\} \right\} $
to the domain $\left\{ 0,\dots,q-1\right\} $. In the following, we
will show explicitly how to do this for the binary case $q=2$, but
the argument can be easily generalized to arbitrary $q$. Note that,
while in the binary case the $\Psi$ functions are linear and thus
trivially concave, in the general case we need to ensure that both
the initial values $\Phi_{i\to\mu}^{0}\left(W_{i}\right)$ and the
external fields $\Gamma_i\left(W_{i}\right)$ are concave; in such case,
the iteration equations \eqref{eq:MS_CavPhi}, \eqref{eq:RMS_CavPsi}
and \eqref{eq:RMS_Psi} ensure that the concavity property holds for
all time steps $t>0$.

\subsection{The binary case}

We will show explicitly how to perform efficiently the computations
for the binary case. In this case we can simplify the notation by parametrizing the message with a single scalar, i.e.~we can write $\Phi_{\mu\to i}^{t}\left(W_{i}\right)=W_{i}\phi_{\mu\to i}^{t}$
and $\Psi_{i\to\mu}^{t}\left(W_{i}\right)=W_{i}\psi_{i\to\mu}^{t}$
and $\Gamma_{i}\left(W_{i}\right)=W_{i}\gamma_{i}$. Eqs.~\eqref{eq:MS_CavPhi}
and \eqref{eq:RMS_CavPsi} then become:
\begin{eqnarray}
\phi_{\mu\to i}^{t+1} & = & \frac{1}{2}\bigg(\max_{\mathbf{W}:\,W_{i}=1\wedge E_{\mu}\left(W\right)=0}\sum_{j\ne i}W_{j}\psi_{j\to\mu}^{t} -\max_{\mathbf{W}:\,W_{i}=-1\wedge E_{\mu}\left(W\right)=0}\sum_{j\ne i}W_{j}\psi_{j\to\mu}^{t}\bigg)\label{eq:MS_Cavphi}\\
\psi_{i\to\mu}^{t} & = & r\,t\,\psi_{i}^{t-1}+\gamma_{i}+\sum_{\mu^{\prime}\ne\mu}\phi_{\mu^{\prime}\to i}^{t}\label{eq:RMS_Cavpsi}
\end{eqnarray}

Correspondingly, eqs.~\eqref{eq:RMS_Psi} and \eqref{eq:MS_Wt} simplify
to:
\begin{eqnarray}
\psi_{i}^{t} & = & r\,t\,\psi_{i}^{t-1}+\gamma_{i}+\sum_{\mu}\phi_{\mu\to i}^{t}\label{eq:RMS_psi}\\
W_{i}^{t} & = & \mathrm{sign}\psi_{i}^{t}
\end{eqnarray}

In order to apply the results of the previous sections, and perform
efficiently the trace over all possible assignments of $W$ of eq.~\eqref{eq:MS_Cavphi},
we first introduce the auxiliary quantities
\begin{eqnarray}
F_{\mu}^{t}\left(\mathbf{S}\right) & = & \sum_{i}S_{i}\xi_{i}^{\mu}\psi_{i\to\mu}^{t}\\
\mathcal{F}_{\mu}^{t}\left(\Delta\right) & = & \max_{\left\{ \mathbf{S}:\,\sum_{i}S_{i}=\Delta\right\} }F_{\mu}^{t}\left(\mathbf{S}\right)
\end{eqnarray}

For simplicity of notation, we will temporarily drop the indices $\mu$
and $t$. We will also assume that all values $\psi_{i\to\mu}^{t}$
are different: as remarked above, the presence of term $\Gamma_{i}$
is sufficient to ensure that this is the case, and otherwise we can
impose an arbitrary order without loss of generality. With this assumption,
the function $\mathcal{F}$, which is defined over $\Delta=\left\{ -N,-N+2,\dots,N-2,N\right\} $,
has a single absolute maximum, and is indeed concave. The absolute
maximum is obtained with the special configuration $\tilde{\mathbf{S}}=\mathrm{argmax}_{\left\{ S_{i}\right\} _{i}}F\left(\mathbf{S}\right)$,
which is trivially obtained by setting $\tilde{S}_{i}=\xi_{i}^{\mu}\mathrm{sign}\psi_{i\to\mu}^{t}$
for all $i$. This configuration corresponds to a value $\tilde{\Delta}=\sum_{i}\tilde{S}_{i}$.
Any variable flip with respect to this configuration, i.e.~any $i$
for which $S_{i}=-\tilde{S}_{i}$, adds a ``cost'' $\Delta F_{i}=2\left|\psi_{i\to\mu}^{t}\right|$
in terms of $F\left(\mathbf{S}\right)$. Therefore, if we partition
the indices $i$ in two groups $S_{+}$ and $S_{-}$ defined by $S_{\pm}=\left\{ i:\,\tilde{S}_{i}=\pm1\right\} $,
and we sort the indices within each group in ascending order according
to $\Delta F_{i}$, we can compute the function $\mathcal{F}\left(\Delta\right)$
for each $\Delta$ by constructively computing the corresponding optimal
configuration $\mathbf{S}$, in the following way: we start from $\mathcal{F}\left(\tilde{\Delta}\right)$,
then proceed in steps of $2$ in both directions subtracting the values
$\Delta F_{i}$ in ascending order, using the variable indices in
$S_{+}$ for $\Delta<\tilde{\Delta}$ and those in $S_{-}$ for $\Delta>\tilde{\Delta}$.

This procedure also associates a ``turning point'' $T_{i}$ to each
index $i$, defined as the value of $\Delta$ for which the optimal
value of $W_{i}$ changes sign, or equivalently such that $\mathcal{F}\left(T_{i}+\tilde{S}_{i}\right)-\mathcal{F}\left(T_{i}-\tilde{S}_{i}\right)=\Delta F_{i}$.
This also implies that: 
\begin{equation}
\mathcal{F}\left(\Delta\right)=\mathcal{F}\left(\tilde{\Delta}\right)-\sum_{i}\Theta\left(\tilde{S}_{i}\left(T_{i}-\Delta\right)\right)\Delta F_{i}\label{eq:Fcal}
\end{equation}
We can also bijectively associate an index to each value of $T_{i}$,
by defining $j_{k}$ such that $T_{j_{k}}=k$.

Next, consider the same quantity where a variable $i$ is left out
of the sum (see eq.~\eqref{eq:convolution})
\begin{eqnarray}
F^{\left(i\right)}\left(\mathbf{S}^{\left(i\right)}\right) & = & \sum_{j\ne i}S_{j}\xi_{j}^{\mu}\psi_{j\to\mu}^{t}\\
\mathcal{F}^{\left(i\right)}\left(\Delta\right) & = & \max_{\left\{ S_{j}\right\} _{j\ne i}:\,\sum_{j\neq i}S_{j}=\Delta}F^{\left(i\right)}\left(\mathbf{S}^{\left(i\right)}\right)\label{eq:cav_Fcal}
\end{eqnarray}
Clearly, one gets the same overall picture as before, except with
a shifted argmax, and shifted turning points. The shifts can be easily
expressed in terms of the previous quantities, and the expressions
used for computing eq.~\eqref{eq:MS_Cavphi} as:
\begin{eqnarray}
\phi_{\mu\to i}^{t+1} & = & \frac{\xi_{i}^{\mu}}{2}\,\left(\max_{\Delta:\,\sigma_{D}^{\mu}\Delta>0}\mathcal{F}^{\left(i\right)}\left(\Delta-1\right)-\max_{\Delta:\,\sigma_{D}^{\mu}\Delta>0}\mathcal{F}^{\left(i\right)}\left(\Delta+1\right)\right)\label{eq:MS_Cavphi_intermediate}
\end{eqnarray}

The full details of the computation are provided in the Appendix,
Sec.~\ref{sub:gory-details}. Here, we report the end result:
\begin{eqnarray}
\phi_{\mu\to i}^{t+1} & = & \xi_{i}^{\mu}\,\left(\Theta\left(\sigma_{D}^{\mu}\right)\Theta\left(-\tilde{\Delta}+\tilde{S}_{i}+1\right)\left(\Theta\left(T_{i}-1\right)h_{j_{0}}+\Theta\left(-T_{i}+1\right)h_{j_{2}}\right)\right.+\label{eq:MS_Cavphi_final}\\
 &  & \:+\left.\Theta\left(-\sigma_{D}^{\mu}\right)\Theta\left(\tilde{\Delta}-\tilde{S}_{i}+1\right)\left(\Theta\left(T_{i}+1\right)h_{j_{-2}}+\Theta\left(-T_{i}-1\right)h_{j_{0}}\right)\right)\nonumber 
\end{eqnarray}
where $h_{j}=-\xi_{j}^{\mu}\psi_{j\to\mu}^{t}$. From this expression,
we see that we can update the cavity fields $\phi_{\mu\to i}$ very
efficiently for all $i$, using the following procedure:
\begin{itemize}
\item We do one pass of the whole array of $h_{i}$ by which we determine
the $\tilde{S}_{i}$ values, we split the indices $j$ into $S_{+}$
and $S_{-}$ and we compute $\tilde{\Delta}$. This requires $\mathcal{O}\left(N\right)$
operations (all of which are trivial).
\item We separately partially sort the indices in $S_{+}$ and $S_{-}$
and get $j_{-2}$, $j_{0}$ and $j_{2}$ and the turning points $T_{i}$.
This requires at most $\mathcal{O}\left(N\log N\right)$ operations.
Note that we can use a partial sort because we computed $\tilde{\Delta}$,
and so we know how many indices we need to sort, and from which set
$S_{\pm}$, until we get to the ones with turning points around $0$;
also, we are only interested in computing $\Theta\left(T_{i}-1\right)$
and $\Theta\left(T_{i}+1\right)$ instead of all values of $T_{i}$.
This makes it likely for the procedure to be significantly less computationally
expensive than the worst case scenario.
\item For each $i$ we compute $\phi_{\mu\to i}^{t+1}$ from the equation
above. This requires $\mathcal{O}\left(1\right)$ operations (implemented
in practice with three conditionals and a lookup).
\end{itemize}

\section{Numerical results\label{sec:Numerical-results}}

We tested extensively the binary case $q=2$ with $W_{i}\in\left\{ -1,+1\right\} $
and the ternary case $q=3$ with $W_{i}\in\left\{ -1,0,1\right\} $,
for single layer networks.

\begin{figure}
\includegraphics[width=1\columnwidth]{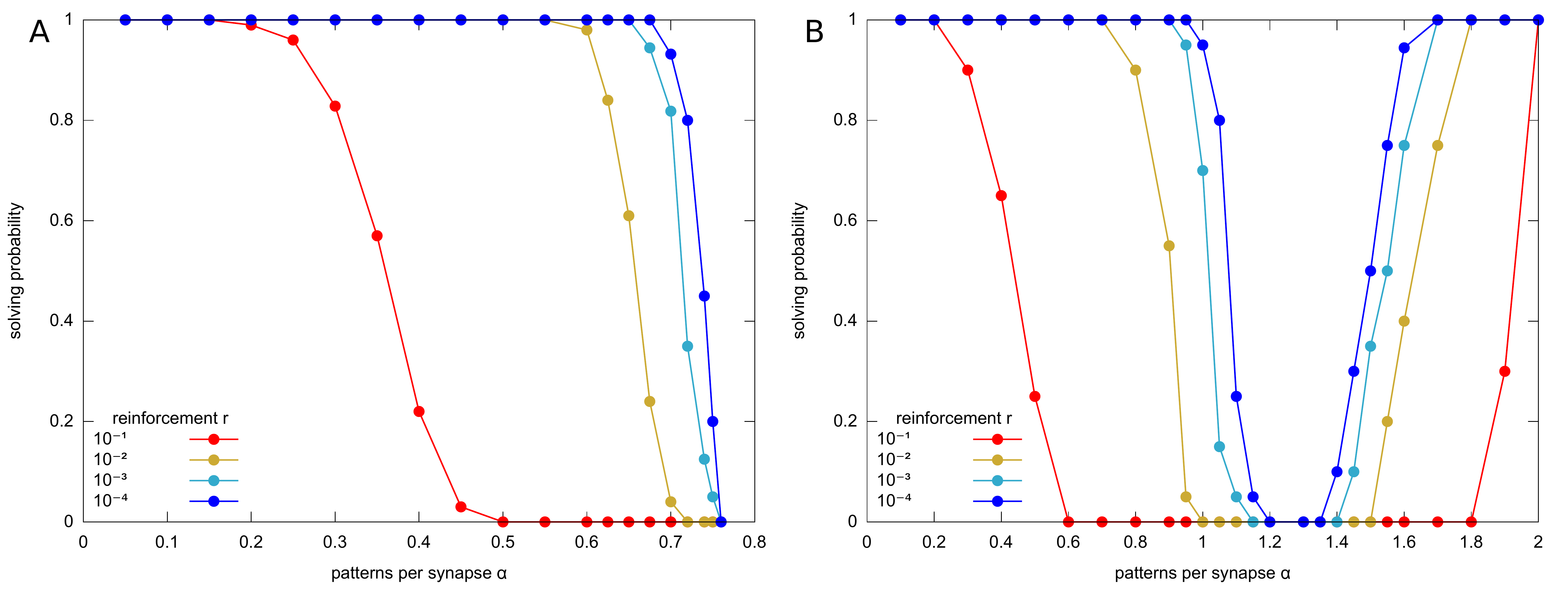}\protect\caption{\textbf{\label{fig:Sol_prob}Solving probability.} Probability of
finding a solution for different values of $\alpha$, in the binary
perceptron case with $N=1001$, with different values of the reinforcement
rate parameter $r$. Performance improves with lower values of $r$.
\textbf{A.} Classification case, $100$ samples per point. The theoretical
capacity is $\alpha_{c}=0.83$ in this case. \textbf{B.} Generalization
case, $20$ samples per point. In this case, the problem has multiple
solutions up to $\alpha_{TS}=1.245$, after which the only solution
is the teacher.}

\end{figure}

We start from the binary case. Fig.~\ref{fig:Sol_prob} shows the
probability of finding a solution when fixing the reinforcement rate
$r$, for different values of $r$ and $\alpha$. Reducing $r$ allows
to reach higher values of $\alpha$; the shape of the curves suggest
that in the limit $r\to0$ there would be sharp transitions at critical
values of $\alpha$'s. In the classification case, Fig.~\ref{fig:Sol_prob}A,
the transition is around $\alpha\simeq0.75$, while the theoretical
critical capacity is $\alpha_{c}=0.83$. This value is comparable
to the one obtained with the reinforced BP algorithm of \cite{braunstein_learning_2006}.
In the generalization case, there are two transitions: the first one
occurs around $\alpha\simeq1.1$, before the first-order transition
at $\alpha_{TS}=1.245$ where, according to the theory, the only solution
is the teacher; the second transition occurs around $\alpha\simeq1.5$.
This second transition is compatible with the end of the meta-stable
regime (see e.g.~\cite{engel_statistical_2001}); indeed, after this
point the algorithm is able to correctly infer the teacher perceptron.

\begin{figure}
\includegraphics[width=1\columnwidth]{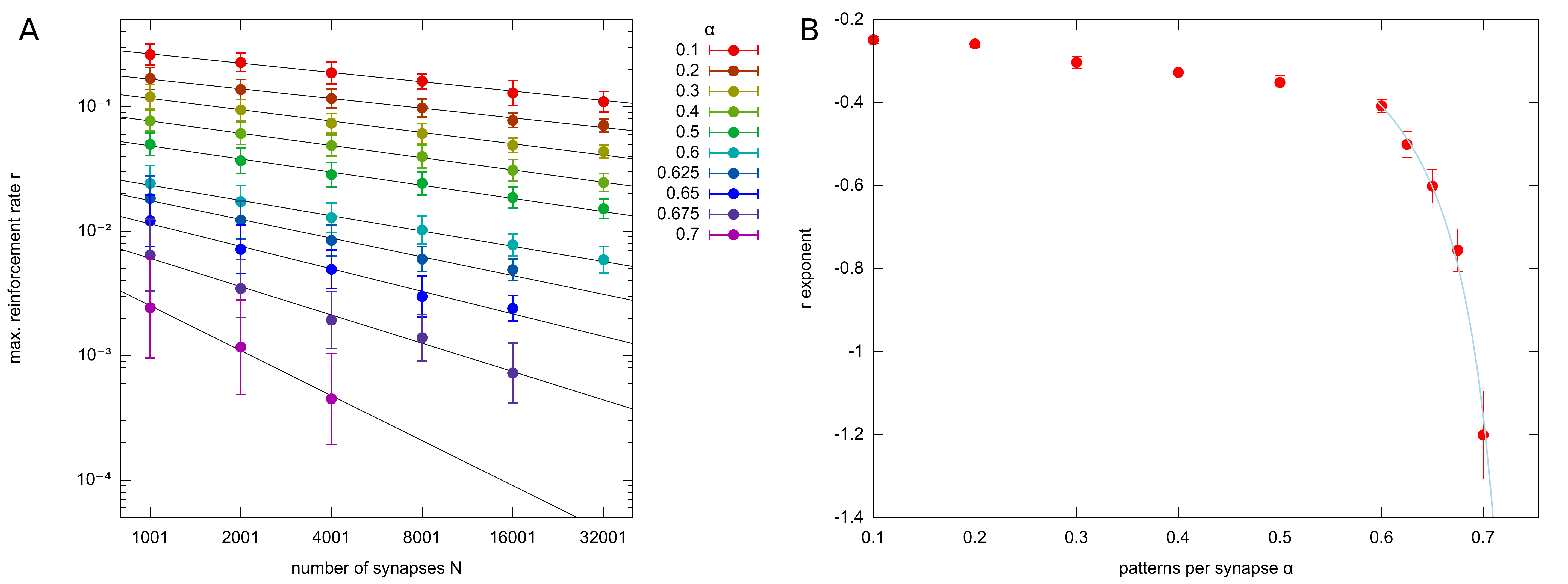}

\protect\caption{\textbf{\label{fig:Max_r}Maximum reinforcement rate.} \textbf{A.}
Average value of the maximum reinforcement rate $r$ which allows to find
a solution, in the binary perceptron classification case, at various
values of $N$ and $\alpha$, in log-log scale. The reinforcement
rate decreases with $N$ and $\alpha$. Error bars show the standard
deviation of the distributions. Black lines show the result of fits
of the form $r\left(\alpha,N\right)=a\left(\alpha\right)N^{b\left(\alpha\right)}$,
one for each value of $\alpha$. The number of samples varied between
$100$ for $N=1001$ and $10$ for $N=32001$. \textbf{B.} The fitted
values of the exponents $b\left(\alpha\right)$ in panel A. The continuous
curve shows a fit of the data for $\alpha\ge0.6$ by the function
$b\left(\alpha\right)=\frac{c}{\alpha_{U}-\alpha}$. The fit yields
$c=-0.063\pm0.002$ and $\alpha_{U}=0.755\pm0.004$. The value of
$\alpha_{U}$ is an estimate of the critical capacity of the algorithm.}
\end{figure}

A second batch of experiments on the same architecture, in the classification
case, is shown in Fig.~\ref{fig:Max_r}. In this case, we estimated
the maximum value of $r$ which allows to find a solution, at different
values of $N$ and $\alpha$; i.e.~for each test sample we started
from a high value of $r$ (e.g.~$r=10^{-1}$) and checked if the
algorithm was able to find a solution; if the algorithm failed, we
reduced $r$ and tried the same sample again. In the cases shown,
the solution was always found eventually. The results indicate that
the value of $r$ required decreases with $N$, and the behaviour
is well described by a power low, i.e.~$r=aN^{b}$ with $a<0$ and
$b<0$, where the values of $a$ and $b$ depend on $\alpha$. Since
the number of iterations required is inversely proportional to $r$
(not shown), this implies that the overall solving time of the MS
algorithm is of $O\left(N^{1-b}\log\left(N\right)\right)$, i.e.~it
is worse than the reinforced BP in this respect. The value of $b$
is between $0$ and $-0.5$ up to $\alpha=0.6$, after which its magnitude
decreases abruptly (see Fig.~\ref{fig:Max_r}B). The behaviour for
large $\alpha$ seems to be reasonably well fit by a curve $b\left(\alpha\right)=\frac{c}{\alpha_{U}-\alpha}$,
suggesting the presence of a vertical asymptote at $\alpha_{U}=0.755\pm0.004$,
which is an estimate of the critical capacity of the algorithm in
the limit of large $N$.

\begin{figure}
\begin{centering}
\includegraphics[width=0.5\textwidth]{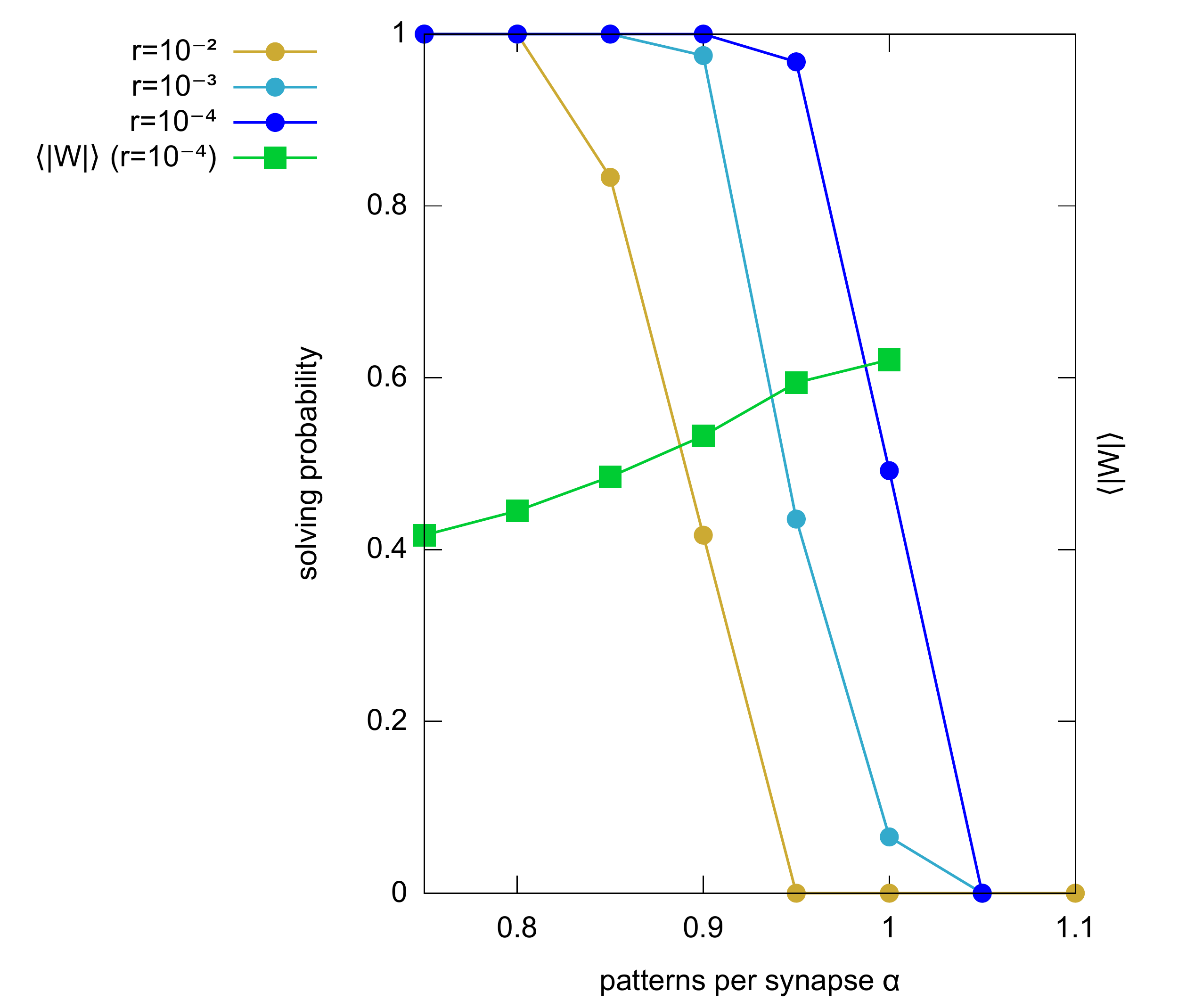}
\par\end{centering}

\protect\caption{\label{fig:3states}Learning of random $\left\{ -1,1\right\} $ patterns with a ternary
$w=\left\{ -1,0,1\right\} $ perceptron, with dilution (regularization)
prior term; $N=1001$, $20$ samples per point. For solved instances
with $r=10^{-4}$, the average fraction of non-zero weights is also
shown (standard deviations smaller than point size).}
\end{figure}

In the ternary single layer case, we tested learning of $\left\{ -1,1\right\} $
random patterns with ternary $\left\{ -1,0,1\right\} $ weights and
concave bias (i.e. prior). In practice, we use the function $\Gamma_{i}\left(W_{i}\right)=\lambda\delta\left(W_{i}\right)+\Gamma'_{i}\left(W_{i}\right)$
(where $\Gamma'$ is the symmetry-breaking noise term and $\lambda$ is
sufficiently large) to favour zero weights, so solutions with a minimimal
number of zeros are searched, i.e. we add an $l_{0}$ regularization
term. Results (See Fig.~\ref{fig:3states}) are qualitatively similar to the $\left\{ -1,1\right\} $
case with a larger capacity (around $\alpha=1$; the critical capacity
is $\alpha_{c}=1.17$ in this case). The average non-zero weights
in a solution grows when getting closer to the critical $\alpha$
up to a value that is smaller than $2/3$ (the value that makes the
entropy of unconstrained $\left\{ -1,0,1\right\} $ perceptrons largest).

In the fully-connected multi-layer case, the algorithm does not get
as close to the critical capacity as for the single-layer case, but
it is still able to achieve non-zero capacity in rather large instances.
For example, in the classification case with binary synapses, $N=1001$
inputs, $K=3$ hidden units, the algorithmic critical capacity is
$\alpha\simeq0.33$ when $r=10^{-5}$ (tested on $20$ samples), corresponding
to storing $M=1001$ patterns with $3003$ weights (thus demonstrating
a greater discriminatory power than the single-layer case with the
same input size). The reason for the increased difficulty in this
case is not completely clear: we speculate that it is both due to
the permutation symmetry between the hidden units and to replica-symmetry-breaking
effects: these effects tend to trap the algorithm --- in its intermediate
steps --- in states which mix different clusters of solutions, making
convergence difficult. Still, the use of symmetry-breaking noise helps
achieving non trivial results even in this case, which constitutes
an improvement with respect to the standard BP algorithm.

\section{Conclusions}

Up to now, the large $N$ limit could be exploited on BP equations
for the learning problem with discrete synapses to obtain an extremely
simple set of approximated equations that made the computation of
an iteration to scale linearly with the problem size $MN$. For the
MS equations however, those approximations cannot be made and a naive
approximation scales as $MN^{3}$ which is normally too slow for most
practical purposes. In this work, we showed that MS equations can
be computed exactly (with no approximations) in time $MN\log N$,
rendering the approach extremely interesting in practice. A word is in order about the MS
equations with reinforcement term, which we propose as a valid alternative to Belief Propagations-based
methods. Although we cannot claim any universal property of the reinforced equations from theoretical arguments and we only tested a limited number of cases,  
extensive simulations for these cases and previous results obtained by applying the same technique to other optimization problems of very different nature \cite{bailly-bechet_finding_2011, anthony_gitter_sharing_2013, altarelli_optimizing_2013, altarelli_containing_2014} have confirmed the same qualitative behaviour; that is, that the number of iterations until convergence scales as $r^{-1}$ and that results monotonically improve as $r$ decreases.  
As an additional advantage of MS, inherent symmetries present in
the original system are naturally broken thanks to ad-hoc noise terms
that are standard in MS. The MS equations are additionally computationally
simpler because they normally require only sum and max operations,
in contrast with hyperbolic trigonometric functions required by BP
equations. Extensive simulations for discrete $\{-1,1\}$ and $\{-1,0,1\}$
weights show that the performance is indeed very good, and the algorithm
achieves a capacity close to the theoretical one (very similar to
the one of Belief Propagation). 
\begin{acknowledgments}
CB acknowledges the European Research Council for grant n\textdegree{}
267915.
\end{acknowledgments}

\section*{Appendix}

\subsection*{Details of the computation of the cavity fields.\label{sub:gory-details}}

In this section we provide the full details of the computation leading
to eq.~\eqref{eq:MS_Cavphi_final}.

As noted in the main text, the expression of the cavity quantities
$\mathcal{F}^{\left(i\right)}\left(\Delta\right)$ (see eq.~\eqref{eq:cav_Fcal})
is analogous to that of the non-cavity counterpart $\mathcal{F}\left(\Delta\right)$
(eq.~\eqref{eq:Fcal}), where the argmax has changed to $\tilde{\Delta}^{\left(i\right)}=\tilde{\Delta}-\tilde{S}_{i}$,
and the turning points have changed: 
\begin{equation}
\mathcal{F}^{\left(i\right)}\left(\Delta\right)=\mathcal{F}^{\left(i\right)}\left(\tilde{\Delta}^{\left(i\right)}\right)-\sum_{j\ne i}\Theta\left(\tilde{S}_{j}\left(T_{j}^{\left(i\right)}-\Delta\right)\right)\Delta F_{j}
\end{equation}

We need to express the relationship between the old turning points
and the new ones: having omitted the variable $i$, it means that
there is a global shift of $-\tilde{S_{i}}$, and that the turning
points to the left (right) of $T_{i}$ have shifted to the right (left)
if $\tilde{S}_{i}=1$ ($\tilde{S}_{i}=-1$, respectively):
\begin{eqnarray}
T_{j}^{\left(i\right)} & = & T_{j}-\tilde{S}_{i}+2\tilde{S}_{i}\Theta\left(\tilde{S}_{i}\left(T_{i}-T_{j}\right)\right)\nonumber \\
 & = & T_{j}+\mbox{sign}\left(T_{i}-T_{j}-\tilde{S}_{i}\right)
\end{eqnarray}
(note that we chose to use the convention that $\Theta\left(0\right)=0$).

Therefore we obtain:
\begin{equation}
\mathcal{F}^{\left(i\right)}\left(\Delta\right)=\mathcal{F}^{\left(i\right)}\left(\tilde{\Delta}^{\left(i\right)}\right)-\sum_{j\ne i}\Theta\left(\tilde{S}_{j}\left(T_{j}+\mbox{sign}\left(T_{i}-T_{j}-\tilde{S}_{i}\right)-\Delta\right)\right)\Delta F_{j}\label{eq:cav_Fcal-2}
\end{equation}

Next, we consider the cavity quantity:
\begin{eqnarray}
\mathcal{C}_{i}\left(\Delta,S_{i}\right) & = & \max_{\left\{ S_{j}\right\} _{j\ne i}:\,\sum_{j}S_{j}\xi_{j}^{\mu}=\Delta}\left(\sum_{j\ne i}S_{j}\xi_{j}^{\mu}\psi_{j\to\mu}^{t}\right)\nonumber \\
 & = & \mathcal{F}^{\left(i\right)}\left(\Delta-S_{i}\right)
\end{eqnarray}

which allows us to write eq.~\eqref{eq:MS_Cavphi} as
\begin{eqnarray}
\phi_{\mu\to i}^{t+1} & = & \frac{\xi_{i}^{\mu}}{2}\,\left(\max_{\Delta:\,\sigma_{D}^{\mu}\Delta>0}\mathcal{C}_{i}\left(\Delta,\xi_{i}^{\mu}\right)-\max_{\Delta:\,\sigma_{D}^{\mu}\Delta>0}\mathcal{C}_{i}\left(\Delta,-\xi_{i}^{\mu}\right)\right)\nonumber \\
 & = & \frac{\xi_{i}^{\mu}}{2}\,\left(\max_{\Delta:\,\sigma_{D}^{\mu}\Delta>0}\mathcal{F}^{\left(i\right)}\left(\Delta-1\right)-\max_{\Delta:\,\sigma_{D}^{\mu}\Delta>0}\mathcal{F}^{\left(i\right)}\left(\Delta+1\right)\right)
\end{eqnarray}
(this is eq.~\eqref{eq:MS_Cavphi_intermediate} in the main text).

Note that $\mathcal{F}^{\left(i\right)}\left(\Delta\right)$ is concave
and has a maximum at $\tilde{\Delta}^{\left(i\right)}=\tilde{\Delta}-\tilde{S}_{i}$.
Using this fact, and eq.~\eqref{eq:cav_Fcal-2}, we can derive explicit
formulas for the expressions which appear in the cavity field, by
considering the two cases for $\sigma_{D}^{\mu}$ separately, and
simplifying the result with simple algebraic manipulations afterwards:
\begin{eqnarray*}
\max_{\Delta:\,\sigma_{D}^{\mu}\Delta>0}\mathcal{F}^{\left(i\right)}\left(\Delta-1\right) & = & \Theta\left(\sigma_{D}^{\mu}\right)\left(\Theta\left(\tilde{\Delta}-\tilde{S}_{i}+1\right)\mathcal{F}^{\left(i\right)}\left(\tilde{\Delta}-\tilde{S}_{i}\right)+\Theta\left(-\tilde{\Delta}+\tilde{S}_{i}-1\right)\mathcal{F}^{\left(i\right)}\left(0\right)\right)+\\
 &  & +\Theta\left(-\sigma_{D}^{\mu}\right)\left(\Theta\left(-\tilde{\Delta}+\tilde{S}_{i}-1\right)\mathcal{F}^{\left(i\right)}\left(\tilde{\Delta}-\tilde{S}_{i}\right)+\Theta\left(\tilde{\Delta}-\tilde{S}_{i}+1\right)\mathcal{F}^{\left(i\right)}\left(-2\right)\right)\\
 & = & \Theta\left(\sigma_{D}^{\mu}\left(\tilde{\Delta}-\tilde{S}_{i}+1\right)\right)\mathcal{F}^{\left(i\right)}\left(\tilde{\Delta}-\tilde{S}_{i}\right)+\\
 &  & +\Theta\left(\sigma_{D}^{\mu}\right)\Theta\left(-\tilde{\Delta}+\tilde{S}_{i}-1\right)\mathcal{F}^{\left(i\right)}\left(0\right)+\\
 &  & +\Theta\left(-\sigma_{D}^{\mu}\right)\Theta\left(\tilde{\Delta}-\tilde{S}_{i}+1\right)\mathcal{F}^{\left(i\right)}\left(-2\right)
\end{eqnarray*}

\begin{eqnarray*}
\max_{\Delta:\,\sigma_{D}^{\mu}\Delta>0}\mathcal{F}^{\left(i\right)}\left(\Delta+1\right) & = & \Theta\left(\sigma_{D}^{\mu}\right)\left(\Theta\left(\tilde{\Delta}-\tilde{S}_{i}-1\right)\mathcal{F}^{\left(i\right)}\left(\tilde{\Delta}-\tilde{S}_{i}\right)+\Theta\left(-\tilde{\Delta}+\tilde{S}_{i}+1\right)\mathcal{F}^{\left(i\right)}\left(2\right)\right)+\\
 &  & +\Theta\left(-\sigma_{D}^{\mu}\right)\left(\Theta\left(-\tilde{\Delta}+\tilde{S}_{i}+1\right)\mathcal{F}^{\left(i\right)}\left(\tilde{\Delta}-\tilde{S}_{i}\right)+\Theta\left(\tilde{\Delta}-\tilde{S}_{i}-1\right)\mathcal{F}^{\left(i\right)}\left(0\right)\right)\\
 & = & \Theta\left(\sigma_{D}^{\mu}\left(\tilde{\Delta}-\tilde{S}_{i}-1\right)\right)\mathcal{F}^{\left(i\right)}\left(\tilde{\Delta}-\tilde{S}_{i}\right)+\\
 &  & +\Theta\left(\sigma_{D}^{\mu}\right)\Theta\left(-\tilde{\Delta}+\tilde{S}_{i}+1\right)\mathcal{F}^{\left(i\right)}\left(2\right)+\\
 &  & +\Theta\left(-\sigma_{D}^{\mu}\right)\Theta\left(\tilde{\Delta}-\tilde{S}_{i}-1\right)\mathcal{F}^{\left(i\right)}\left(0\right)
\end{eqnarray*}

Plugging these back in the expression for the cavity field, we can
reach --- again by simple algebraic manipulations --- an expression
which only uses $\mathcal{F}^{\left(i\right)}\left(-2\right)$, $\mathcal{F}^{\left(i\right)}\left(0\right)$
and $\mathcal{F}^{\left(i\right)}\left(2\right)$:

\begin{eqnarray*}
\phi_{\mu\to i}^{t+1} & = & \frac{\xi_{i}^{\mu}}{2}\,\left(\left(\Theta\left(\sigma_{D}^{\mu}\left(\tilde{\Delta}-\tilde{S}_{i}+1\right)\right)-\Theta\left(\sigma_{D}^{\mu}\left(\tilde{\Delta}-\tilde{S}_{i}-1\right)\right)\right)\mathcal{F}^{\left(i\right)}\left(\tilde{\Delta}-\tilde{S}_{i}\right)\right.+\\
 &  & \qquad+\Theta\left(\sigma_{D}^{\mu}\right)\left(\Theta\left(-\tilde{\Delta}+\tilde{S}_{i}-1\right)\mathcal{F}^{\left(i\right)}\left(0\right)-\Theta\left(-\tilde{\Delta}+\tilde{S}_{i}+1\right)\mathcal{F}^{\left(i\right)}\left(2\right)\right)+\\
 &  & \qquad+\left.\Theta\left(-\sigma_{D}^{\mu}\right)\left(\Theta\left(\tilde{\Delta}-\tilde{S}_{i}+1\right)\mathcal{F}^{\left(i\right)}\left(-2\right)-\Theta\left(\tilde{\Delta}-\tilde{S}_{i}-1\right)\mathcal{F}^{\left(i\right)}\left(0\right)\right)\right)\\
 & = & \frac{\xi_{i}^{\mu}}{2}\,\left(\delta\left(\tilde{\Delta},\tilde{S}_{i}\right)\mbox{sign}\left(\sigma_{D}^{\mu}\right)\mathcal{F}^{\left(i\right)}\left(0\right)\right.+\\
 &  & \qquad+\Theta\left(\sigma_{D}^{\mu}\right)\left(-\delta\left(\tilde{\Delta},\tilde{S}_{i}\right)\mathcal{F}^{\left(i\right)}\left(2\right)+\Theta\left(-\tilde{\Delta}+\tilde{S}_{i}-1\right)\left(\mathcal{F}^{\left(i\right)}\left(0\right)-\mathcal{F}^{\left(i\right)}\left(2\right)\right)\right)+\\
 &  & \qquad+\left.\Theta\left(-\sigma_{D}^{\mu}\right)\left(\delta\left(\tilde{\Delta},\tilde{S}_{i}\right)\mathcal{F}^{\left(i\right)}\left(-2\right)+\Theta\left(\tilde{\Delta}-\tilde{S}_{i}-1\right)\left(\mathcal{F}^{\left(i\right)}\left(-2\right)-\mathcal{F}^{\left(i\right)}\left(0\right)\right)\right)\right)\\
 & = & \frac{\xi_{i}^{\mu}}{2}\,\left(\delta\left(\tilde{\Delta},\tilde{W}_{i}\xi_{i}^{\mu}\right)\left(\mbox{sign}\left(\sigma_{D}^{\mu}\right)\left(\mathcal{F}^{\left(i\right)}\left(0\right)-\mathcal{F}^{\left(i\right)}\left(2\sigma_{D}^{\mu}\right)\right)\right)\right.+\\
 &  & \qquad+\Theta\left(\sigma_{D}^{\mu}\right)\Theta\left(-\tilde{\Delta}+\tilde{W}_{i}\xi_{i}^{\mu}-1\right)\left(\mathcal{F}^{\left(i\right)}\left(0\right)-\mathcal{F}^{\left(i\right)}\left(2\right)\right)+\\
 &  & \qquad+\left.\Theta\left(-\sigma_{D}^{\mu}\right)\Theta\left(\tilde{\Delta}-\tilde{W}_{i}\xi_{i}^{\mu}-1\right)\left(\mathcal{F}^{\left(i\right)}\left(-2\right)-\mathcal{F}^{\left(i\right)}\left(0\right)\right)\right)\\
 & = & \frac{\xi_{i}^{\mu}}{2}\,\left(\Theta\left(\sigma_{D}^{\mu}\right)\Theta\left(-\tilde{\Delta}+\tilde{S}_{i}+1\right)\left(\mathcal{F}^{\left(i\right)}\left(0\right)-\mathcal{F}^{\left(i\right)}\left(2\right)\right)\right.+\\
 &  & \:+\left.\Theta\left(-\sigma_{D}^{\mu}\right)\Theta\left(\tilde{\Delta}-\tilde{S}_{i}+1\right)\left(\mathcal{F}^{\left(i\right)}\left(-2\right)-\mathcal{F}^{\left(i\right)}\left(0\right)\right)\right)
\end{eqnarray*}

These expressions can be further simplified, since the differences
between the values of $\mathcal{F}^{\left(i\right)}$ at neighboring
values only depends on the ``steps'' induced by the spins which
are associated with turning points in that region:

\begin{eqnarray*}
\mathcal{F}^{\left(i\right)}\left(0\right)-\mathcal{F}^{\left(i\right)}\left(2\right) & = & \sum_{j\ne i}\left(\Theta\left(\tilde{S}_{j}\left(T_{j}+\mbox{sign}\left(T_{i}-T_{j}-\tilde{S}_{i}\right)-2\right)\right)-\Theta\left(\tilde{S}_{j}\left(T_{j}+\mbox{sign}\left(T_{i}-T_{j}-\tilde{S}_{i}\right)\right)\right)\right)\Delta F_{j}\\
 & = & -\sum_{j\ne i}\tilde{S}_{j}\delta\left(T_{j},1-\mbox{sign}\left(T_{i}-T_{j}-\tilde{S}_{i}\right)\right)\Delta F_{j}\\
 & = & -\tilde{S}_{j_{0}}\left(1-\delta\left(i,j_{0}\right)\right)\Theta\left(T_{i}-\tilde{S}_{i}\right)\Delta F_{j_{0}}-\tilde{S}_{j_{2}}\left(1-\delta\left(i,j_{2}\right)\right)\Theta\left(-T_{i}+2+\tilde{S}_{i}\right)\Delta F_{j_{2}}\\
 & = & -\tilde{S}_{j_{0}}\Theta\left(T_{i}-1\right)\Delta F_{j_{0}}-\tilde{S}_{j_{2}}\Theta\left(-T_{i}+1\right)\Delta F_{j_{2}}
\end{eqnarray*}
where in the last step we used the Kronecker deltas to get rid for
the differences between the two cases for $\tilde{S}_{i}$. The other
case is very similar:

\begin{eqnarray*}
\mathcal{F}^{\left(i\right)}\left(-2\right)-\mathcal{F}^{\left(i\right)}\left(0\right) & = & \sum_{j\ne i}\left(\Theta\left(\tilde{S}_{j}\left(T_{j}+\mbox{sign}\left(T_{i}-T_{j}-\tilde{S}_{i}\right)\right)\right)-\Theta\left(\tilde{S}_{j}\left(T_{j}+\mbox{sign}\left(T_{i}-T_{j}-\tilde{S}_{i}\right)+2\right)\right)\right)\Delta F_{j}\\
 & = & -\sum_{j\ne i}\tilde{S}_{j}\delta\left(T_{j},-1-\mbox{sign}\left(T_{i}-T_{j}-\tilde{S}_{i}\right)\right)\Delta F_{j}\\
 & = & -\tilde{S}_{j_{-2}}\left(1-\delta\left(i,j_{-2}\right)\right)\Theta\left(T_{i}+2-\tilde{S}_{i}\right)\Delta F_{j_{-2}}-\tilde{S}_{j_{0}}\left(1-\delta\left(i,j_{0}\right)\right)\Theta\left(\tilde{S}_{i}-T_{i}\right)\Delta F_{j_{0}}\\
 & = & -\tilde{S}_{j_{-2}}\Theta\left(T_{i}+1\right)\Delta F_{j_{-2}}-\tilde{S}_{j_{0}}\Theta\left(-1-T_{i}\right)\Delta F_{j_{0}}
\end{eqnarray*}

Going back to the cavity fields, and defining $h_{j}=-\frac{1}{2}\tilde{S}_{j}\Delta F_{j}=-\xi_{j}^{\mu}\psi_{j\to\mu}^{t}$,
we finally get eq.~\eqref{eq:MS_Cavphi_final}:
\begin{eqnarray*}
\phi_{\mu\to i}^{t+1} & = & \xi_{i}^{\mu}\,\left(\Theta\left(\sigma_{D}^{\mu}\right)\Theta\left(-\tilde{\Delta}+\tilde{S}_{i}+1\right)\left(\Theta\left(T_{i}-1\right)h_{j_{0}}+\Theta\left(-T_{i}+1\right)h_{j_{2}}\right)\right.+\\
 &  & \:+\left.\Theta\left(-\sigma_{D}^{\mu}\right)\Theta\left(\tilde{\Delta}-\tilde{S}_{i}+1\right)\left(\Theta\left(T_{i}+1\right)h_{j_{-2}}+\Theta\left(-T_{i}-1\right)h_{j_{0}}\right)\right)
\end{eqnarray*}

\bibliographystyle{unsrt}
\bibliography{percmaxsum2}

\end{document}